\documentclass[10pt,twocolumn,letterpaper]{article}

\usepackage[pagenumbers]{iccv} 

\definecolor{iccvblue}{rgb}{0.21,0.49,0.74}
\usepackage[pagebackref,breaklinks,colorlinks,allcolors=iccvblue]{hyperref}

\def\paperID{*****} 
\def\confName{ICCV}
\def\confYear{2025}

\title{Multi-Source COVID-19 Detection via Variance Risk Extrapolation}

\author{
Runtian Yuan$^1$, Qingqiu Li$^1$, Junlin Hou$^2$, Jilan Xu$^1$, \\
Yuejie Zhang$^1$, Rui Feng$^1$, Hao Chen$^2$\\
\\
$^1$ Fudan University\\
$^2$ The Hong Kong University of Science and Technology\\
}

\begin{document}
\maketitle
\begin{abstract}
We present our solution for the Multi-Source COVID-19 Detection Challenge, which aims to classify chest CT scans into COVID and Non-COVID categories across data collected from four distinct hospitals and medical centers. A major challenge in this task lies in the domain shift caused by variations in imaging protocols, scanners, and patient populations across institutions.
To enhance the cross-domain generalization of our model, we incorporate Variance Risk Extrapolation (VREx) into the training process. VREx encourages the model to maintain consistent performance across multiple source domains by explicitly minimizing the variance of empirical risks across environments. This regularization strategy reduces overfitting to center-specific features and promotes learning of domain-invariant representations.
We further apply Mixup data augmentation to improve generalization and robustness. Mixup interpolates both the inputs and labels of randomly selected pairs of training samples, encouraging the model to behave linearly between examples and enhancing its resilience to noise and limited data.
Our method achieves an average macro F1 score of 0.96 across the four sources on the validation set, demonstrating strong generalization.
\end{abstract}    
\section{Introduction}
\label{sec:intro}
The global outbreak of COVID-19 has presented unprecedented challenges to public health systems worldwide. Rapid and reliable detection of COVID-19~\cite{arsenos2022large,arsenos2023data,gerogiannis2024covid,kollias2018deep,kollias2020deep,kollias2020transparent,kollias2021mia,kollias2022ai,kollias2023ai,kollias2023deep,kollias2024domain,kollias2024sam2clip2sam} is crucial for timely diagnosis, effective isolation, and treatment of infected individuals. Among available diagnostic tools, chest computed tomography (CT) has emerged as a valuable modality due to its ability to capture early pulmonary manifestations of the disease, including ground-glass opacities and bilateral infiltrates.

Despite its clinical utility, developing robust and generalizable automated detection systems based on CT scans remains a significant challenge—particularly in real-world, multi-institutional settings. CT data collected from different hospitals and medical centers often exhibit substantial variability in terms of acquisition parameters, scanner models, and patient demographics. This domain shift can cause conventional deep learning models, trained under empirical risk minimization (ERM), to overfit to domain-specific artifacts and degrade significantly when evaluated on unseen domains.

To address this issue, we participate in the Multi-Source COVID-19 Detection Challenge, which targets binary classification of COVID / Non-COVID cases using CT scans collected from four diverse medical centers. To improve domain robustness, we adopt Variance Risk Extrapolation (VREx)~\cite{krueger2021vrex} as a key component of our training strategy. VREx aims to minimize the variance of prediction risks across different source domains, thereby encouraging the model to learn domain-invariant features. Additionally, we employ Mixup data augmentation, which linearly interpolates both inputs and labels during training, further enhancing generalization and mitigating overfitting in low-data regimes.

Through this combination of domain-generalization and data augmentation techniques, our method achieves strong performance on a held-out multi-center validation set, reaching an average macro F1-score of 0.96 across the four sources. These results highlight the potential of  domain-aware training strategies for improving the real-world applicability of AI-based COVID-19 detection systems.

\begin{figure*}
    \centering
    \includegraphics[width=1\linewidth]{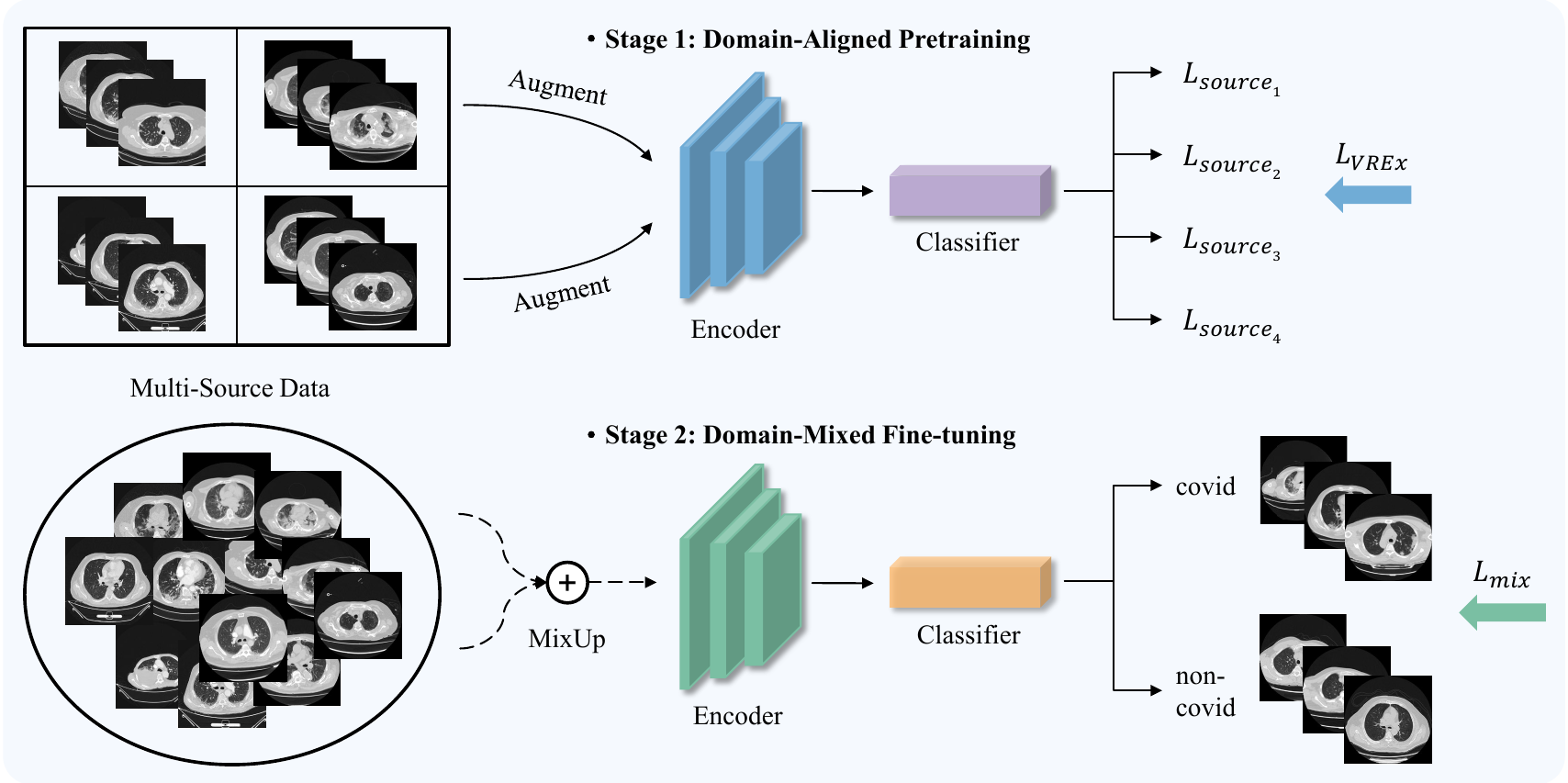}
    \caption{Overview of our framework for multi-source COVID-19 detection.}
    \label{fig:arch}
\end{figure*}

\section{Methodology}

In this paper, we present our framework for the Multi-Source Covid-19 Detection Challenge of PHAROS-AFE-AIMI. Figure~\ref{fig:arch} illustrates the overall architecture of our proposed method, which consists of two key stages: \textbf{Domain-Aligned Pretraining} and \textbf{Domain-Mixed Fine-tuning}. This two-stage framework is designed to fully leverage the diversity of data collected from different sources, which builds upon the foundation laid by our previous work with the CMC network~\cite{hou2021cmc,hou2022cmc_v2,hou2021periphery,hou2022boosting}.

\paragraph{Stage 1: Domain-Aligned Pretraining}

In the pretraining stage, we adopt the VREx loss to align the data distributions from four source domains, each corresponding to a distinct medical center or hospital. The core idea of VREx is to minimize the variance of empirical risk across source domains, encouraging the model to learn domain-invariant representations that generalize well. Specifically, given $n$ source domains, the objective function is defined as:

\begin{equation}
\min_\theta \left( \frac{1}{n} \sum_{i=1}^{n} \mathcal{L}_i(\theta) + \lambda \cdot \mathrm{Var}\left[\mathcal{L}_i(\theta)\right] \right),
\end{equation}
where $\mathcal{L}_i(\theta)$ denotes the loss on the $i$-th domain, and $\lambda$ is a regularization coefficient balancing the empirical risk and its variance. This optimization ensures that the model learns feature representations that perform consistently across domains, providing a robust initialization for the subsequent training stage.

\paragraph{Stage 2: Domain-Mixed Fine-tuning}

Following pretraining, we apply a MixUp-based fine-tuning strategy to further improve generalization. MixUp~\cite{zhang2017mixup} is a data augmentation technique that constructs virtual training samples by linearly interpolating between pairs of examples, potentially from different domains. 

Formally, given two examples $(x_i, y_i)$ and $(x_j, y_j)$, MixUp generates a new sample as:

\begin{equation}
\tilde{x} = \lambda x_i + (1 - \lambda) x_j,\quad
\tilde{y} = \lambda y_i + (1 - \lambda) y_j.
\end{equation}
This cross-domain interpolation encourages the model to learn smoother decision boundaries and reduces overfitting to domain-specific features.

By combining domain alignment and domain mixing, our approach effectively captures invariant features while enhancing generalization to unseen domains. 
Experimental results demonstrate that the proposed method achieves superior performance in COVID-19 detection tasks with improved accuracy and robustness.

\section{Dataset}

\begin{table}[b]
\centering
\begin{tabular}{lccc}
\toprule
\textbf{Set} & \textbf{Total} & \textbf{Non-COVID} & \textbf{COVID} \\
\midrule
Training     & 1124           & 660                & 564            \\
Validation   & 308            & 180                & 128            \\
\bottomrule
\end{tabular}
\caption{Statistics of the CT dataset.}
\label{tab:dataset}
\end{table}

We conduct binary classification of COVID and Non-COVID cases based on chest CT scans collected from four different hospitals and medical centers, labeled as Domains 0 through 3. Each domain represents a distinct source of real-world data, contributing to diversity in imaging protocols and patient demographics.

The training set consists of 1,124 CT scans, including 660 Non-COVID and 564 COVID cases. The validation set includes 308 CT scans, comprising 180 Non-COVID and 128 COVID cases, as summarized in Table~\ref{tab:dataset}.

\section{Experiments}

We evaluate our method on the validation set to assess its effectiveness in distinguishing between COVID and Non-COVID cases. 
Our method achieves an average macro F1 score of \textbf{0.96} on the validation set, demonstrating high accuracy and robustness in COVID-19 detection. 
This result highlights the effectiveness of our domain-aligned and domain-mixed training strategy in learning generalizable representations across institutions.

\section{Conclusion}
In this paper, we propose a two-stage framework for robust COVID-19 detection from chest CT scans, combining domain-aligned pretraining via VREx and domain-mixed fine-tuning with Mixup. By leveraging CT data from four distinct medical institutions, our method effectively addresses domain shift and improves generalization.
Experiments on the validation set demonstrate that our approach achieves an F1 score of 0.96, indicating strong performance in distinguishing COVID from Non-COVID cases. 

{
    \small
    \bibliographystyle{ieeenat_fullname}
    \bibliography{main}

\begin{thebibliography}{18}
\providecommand{\natexlab}[1]{#1}
\providecommand{\url}[1]{\texttt{#1}}
\expandafter\ifx\csname urlstyle\endcsname\relax
  \providecommand{\doi}[1]{doi: #1}\else
  \providecommand{\doi}{doi: \begingroup \urlstyle{rm}\Url}\fi

\bibitem[Arsenos et~al.(2022)Arsenos, Kollias, and Kollias]{arsenos2022large}
Anastasios Arsenos, Dimitrios Kollias, and Stefanos Kollias.
\newblock A large imaging database and novel deep neural architecture for covid-19 diagnosis.
\newblock In \emph{2022 IEEE 14th Image, Video, and Multidimensional Signal Processing Workshop (IVMSP)}, page 1–5. IEEE, 2022.

\bibitem[Arsenos et~al.(2023)Arsenos, Davidhi, Kollias, Prassopoulos, and Kollias]{arsenos2023data}
Anastasios Arsenos, Andjoli Davidhi, Dimitrios Kollias, Panos Prassopoulos, and Stefanos Kollias.
\newblock Data-driven covid-19 detection through medical imaging.
\newblock In \emph{2023 IEEE International Conference on Acoustics, Speech, and Signal Processing Workshops (ICASSPW)}, page 1–5. IEEE, 2023.

\bibitem[Gerogiannis et~al.(2024)Gerogiannis, Arsenos, Kollias, Nikitopoulos, and Kollias]{gerogiannis2024covid}
Demetris Gerogiannis, Anastasios Arsenos, Dimitrios Kollias, Dimitris Nikitopoulos, and Stefanos Kollias.
\newblock Covid-19 computer-aided diagnosis through ai-assisted ct imaging analysis: Deploying a medical ai system.
\newblock In \emph{2024 IEEE International Symposium on Biomedical Imaging (ISBI)}, pages 1--4. IEEE, 2024.

\bibitem[Hou et~al.(2021{\natexlab{a}})Hou, Xu, Feng, Zhang, Shan, and Shi]{hou2021cmc}
Junlin Hou, Jilan Xu, Rui Feng, Yuejie Zhang, Fei Shan, and Weiya Shi.
\newblock Cmc-cov19d: Contrastive mixup classification for covid-19 diagnosis.
\newblock In \emph{Proceedings of the IEEE/CVF International Conference on Computer Vision}, pages 454--461, 2021{\natexlab{a}}.

\bibitem[Hou et~al.(2021{\natexlab{b}})Hou, Xu, Jiang, Du, Feng, Zhang, Shan, and Xue]{hou2021periphery}
Junlin Hou, Jilan Xu, Longquan Jiang, Shanshan Du, Rui Feng, Yuejie Zhang, Fei Shan, and Xiangyang Xue.
\newblock Periphery-aware covid-19 diagnosis with contrastive representation enhancement.
\newblock \emph{Pattern Recognition}, 118:\penalty0 108005, 2021{\natexlab{b}}.

\bibitem[Hou et~al.(2022{\natexlab{a}})Hou, Xu, Zhang, Wang, Zhang, Zhang, and Feng]{hou2022cmc_v2}
Junlin Hou, Jilan Xu, Nan Zhang, Yi Wang, Yuejie Zhang, Xiaobo Zhang, and Rui Feng.
\newblock Cmc\_v2: Towards more accurate covid-19 detection with discriminative video priors.
\newblock In \emph{European Conference on Computer Vision}, pages 485--499. Springer, 2022{\natexlab{a}}.

\bibitem[Hou et~al.(2022{\natexlab{b}})Hou, Xu, Zhang, Zhang, Zhang, and Feng]{hou2022boosting}
Junlin Hou, Jilan Xu, Nan Zhang, Yuejie Zhang, Xiaobo Zhang, and Rui Feng.
\newblock Boosting covid-19 severity detection with infection-aware contrastive mixup classification.
\newblock In \emph{European Conference on Computer Vision}, pages 537--551. Springer, 2022{\natexlab{b}}.

\bibitem[Kollias et~al.(2018)Kollias, Tagaris, Stafylopatis, Kollias, and Tagaris]{kollias2018deep}
Dimitrios Kollias, Athanasios Tagaris, Andreas Stafylopatis, Stefanos Kollias, and Georgios Tagaris.
\newblock Deep neural architectures for prediction in healthcare.
\newblock \emph{Complex \& Intelligent Systems}, 4\penalty0 (2):\penalty0 119–131, 2018.

\bibitem[Kollias et~al.(2020{\natexlab{a}})Kollias, Bouas, Vlaxos, Brillakis, Seferis, Kollia, Sukissian, Wingate, and Kollias]{kollias2020deep}
Dimitrios Kollias, N Bouas, Y Vlaxos, V Brillakis, M Seferis, Ilianna Kollia, Levon Sukissian, James Wingate, and S Kollias.
\newblock Deep transparent prediction through latent representation analysis.
\newblock \emph{arXiv preprint arXiv:2009.07044}, 2020{\natexlab{a}}.

\bibitem[Kollias et~al.(2020{\natexlab{b}})Kollias, Vlaxos, Seferis, Kollia, Sukissian, Wingate, and Kollias]{kollias2020transparent}
Dimitris Kollias, Y Vlaxos, M Seferis, Ilianna Kollia, Levon Sukissian, James Wingate, and Stefanos~D Kollias.
\newblock Transparent adaptation in deep medical image diagnosis.
\newblock In \emph{TAILOR}, page 251–267, 2020{\natexlab{b}}.

\bibitem[Kollias et~al.(2021)Kollias, Arsenos, Soukissian, and Kollias]{kollias2021mia}
Dimitrios Kollias, Anastasios Arsenos, Levon Soukissian, and Stefanos Kollias.
\newblock Mia-cov19d: Covid-19 detection through 3-d chest ct image analysis.
\newblock In \emph{Proceedings of the IEEE/CVF International Conference on Computer Vision}, page 537–544, 2021.

\bibitem[Kollias et~al.(2022)Kollias, Arsenos, and Kollias]{kollias2022ai}
Dimitrios Kollias, Anastasios Arsenos, and Stefanos Kollias.
\newblock Ai-mia: Covid-19 detection and severity analysis through medical imaging.
\newblock In \emph{European Conference on Computer Vision}, page 677–690. Springer, 2022.

\bibitem[Kollias et~al.(2023{\natexlab{a}})Kollias, Arsenos, and Kollias]{kollias2023ai}
Dimitrios Kollias, Anastasios Arsenos, and Stefanos Kollias.
\newblock Ai-enabled analysis of 3-d ct scans for diagnosis of covid-19 \& its severity.
\newblock In \emph{2023 IEEE International Conference on Acoustics, Speech, and Signal Processing Workshops (ICASSPW)}, page 1–5. IEEE, 2023{\natexlab{a}}.

\bibitem[Kollias et~al.(2023{\natexlab{b}})Kollias, Arsenos, and Kollias]{kollias2023deep}
Dimitrios Kollias, Anastasios Arsenos, and Stefanos Kollias.
\newblock A deep neural architecture for harmonizing 3-d input data analysis and decision making in medical imaging.
\newblock \emph{Neurocomputing}, 542:\penalty0 126244, 2023{\natexlab{b}}.

\bibitem[Kollias et~al.(2024{\natexlab{a}})Kollias, Arsenos, and Kollias]{kollias2024domain}
Dimitrios Kollias, Anastasios Arsenos, and Stefanos Kollias.
\newblock Domain adaptation explainability \& fairness in ai for medical image analysis: Diagnosis of covid-19 based on 3-d chest ct-scans.
\newblock In \emph{Proceedings of the IEEE/CVF Conference on Computer Vision and Pattern Recognition}, pages 4907--4914, 2024{\natexlab{a}}.

\bibitem[Kollias et~al.(2024{\natexlab{b}})Kollias, Arsenos, Wingate, and Kollias]{kollias2024sam2clip2sam}
Dimitrios Kollias, Anastasios Arsenos, James Wingate, and Stefanos Kollias.
\newblock Sam2clip2sam: Vision language model for segmentation of 3d ct scans for covid-19 detection.
\newblock \emph{arXiv preprint arXiv:2407.15728}, 2024{\natexlab{b}}.

\bibitem[Krueger et~al.(2021)Krueger, Caballero, Jacobsen, Zhang, Binas, Zhang, Le~Priol, and Courville]{krueger2021vrex}
David Krueger, Ethan Caballero, Joern-Henrik Jacobsen, Amy Zhang, Jonathan Binas, Dinghuai Zhang, Remi Le~Priol, and Aaron Courville.
\newblock Out-of-distribution generalization via risk extrapolation (rex).
\newblock In \emph{International conference on machine learning}, pages 5815--5826. PMLR, 2021.

\bibitem[Zhang et~al.(2017)Zhang, Cisse, Dauphin, and Lopez-Paz]{zhang2017mixup}
Hongyi Zhang, Moustapha Cisse, Yann~N Dauphin, and David Lopez-Paz.
\newblock mixup: Beyond empirical risk minimization.
\newblock \emph{arXiv preprint arXiv:1710.09412}, 2017.

\end{thebibliography}
}

\end{document}